\documentclass[twocolumn,preprintnumbers,amsmath,amssymb,10pt]{revtex4}
\usepackage{amsmath}
\usepackage{graphicx}
\usepackage{epsfig}
\usepackage{epstopdf}
\usepackage{amssymb}
\usepackage{mathrsfs}
\usepackage{bm}
\usepackage{color}
\usepackage{amsmath}
\usepackage{mathrsfs}
\usepackage{algorithm}
\usepackage{algorithmic}
\usepackage{mathtools}
\usepackage{graphicx}
\usepackage{subfigure}
\usepackage{float}
\usepackage{amsthm}
\usepackage{hyperref}
\usepackage{threeparttable}

\begin{document}

\title{Unified monogamy relation of entanglement measures}

\author{Xue Yang and Ming-Xing Luo}

\affiliation{The School of Information Science and Technology, Southwest Jiaotong University, Chengdu 610031, China}

\begin{abstract}
The monogamy of quantum entanglement captures the property of limitation in the distribution of entanglement. Various monogamy relations exist for different entanglement measures that are important in quantum information processing. Our goal in this work is to propose a general monogamy inequality for all entanglement measures on entangled qubit systems. The present result provide a unified model for various entanglement measures including the concurrence, the negativity, the entanglement of formation, Tsallis-$q$ entropy, R\'{e}nyi-$q$ entropy, and Unified-$(q, s)$ entropy. We then proposed tightened monogamy inequalities for multipartite systems. We finally prove a generic result for the tangle of high-dimensional entangled states to show the distinct feature going beyond qubit systems. These results are useful for exploring the entanglement theory, quantum information processing and secure quantum communication.
\end{abstract}

\maketitle

\section{Introduction}

Entanglement as one of the most remarkable phenomena in quantum mechanics reveals the fundamental insights into the nature of quantum correlations. It also provides a crucial resource for quantum information processing, including quantum teleportation \cite{Bennett1}, quantum dense coding \cite{Bennett2}, quantum secret sharing \cite{Hillery}, and quantum cryptography \cite{Gisin}. An important problem is how to explore useful entanglement measures to quantify the entangled systems \cite{Vedral(1997)}.

So far, there are lots of interesting entanglement measures for bipartite entangled systems. The concurrence of bipartite entangled systems is firstly defined by Hill and Wootters \cite{Hill(1997)} using the linear entropy.
Based on von Neumann entropy, the entanglement of formation (EOF) is used to quantify the cost for generating entangled systems by local operations and classical communication (LOCC) \cite{Bennett(1996)3824}. Another well-known measure of entanglement is the negativity \cite{Vidal(2002)} which is based on the positive partial transposition (PPT) criterion \cite{Peres}. As the generalizations of von Neumann entropy, Tsallis-$q$ entropy \cite{LV(1998),Kim(2010)T} and R\'{e}nyi-$q$ entropy \cite{HHH(1996),Gour(2007),Kim(2010)R} are used for quantifying the uncertainty of quantum systems. These one-parameter entanglement measures are then extended to two-parameter entropy, Unified-$(q, s)$ entropy \cite{KimBarry(2011)}.

One of important issues related to the entanglement measure is the limited shareability of bipartite entanglement for multipartite entangled qubit systems which is known as the monogamy of entanglement (MOE)\cite{Terhal(2004)}. It can be displayed as the following inequality:
\begin{eqnarray}
E_{A|BC}\geq E_{A|B}+E_{A|C}
\label{eqn0}
\end{eqnarray}
where $E_{A|BC}(\cdot{})$ is an entanglement measure of a composite quantum system consisting of qubit systems $A$, $B$, and $C$ with respect to the bipartition $A$ and $\{B,C\}$, $E_{A|B}$ and $E_{A|C}$ are entanglement measures of bipartite systems. From the inequality (\ref{eqn0}) there exists a mutually exclusive relation of the bipartite entangled systems between $A$ and each of $B$ and $C$, so that the summation of bipartite entanglement measures is upper bounded by the total quantity between $A$ and $BC$.  Consequently, any measure of bipartite entanglement that does satisfy the inequality (\ref{eqn0}) is called entanglement monogamous \cite{Terhal(2004)}.

Generally, the inequality (\ref{eqn0}) does not hold for all entanglement measures. Thus a natural question is to determine whether a given entanglement measure is monogamous or not. A remarkable result that the square of the concurrence satisfies the monogamy inequality (\ref{eqn0}) was established by Coffman, Kundu and Wootters (CKW) \cite{V. Coffman} for three qubits.  The so-called CKW inequality was later extended to arbitrary $N$-qubit systems \cite{T.J.Osborne}. Interestingly,
it was further proven that similar multiqubit monogamy inequalities can be established for the squared  negativity, the squared convex-roof extended negativity \cite{Kim(2009)N}, the square of EOF \cite{Oliveira(2014),Bai3,Bai(2014)}, Tsallis-$q$ entropy \cite{Kim(2010)T}, the squared Tsallis-$q$ entropy \cite{Luo(2016)}, R\'{e}nyi-$q$ entropy \cite{Kim(2010)R}, the squared R\'{e}nyi-$q$ entropy \cite{R(2015)}, Unified-$(q, s)$ entropy \cite{KimBarry(2011)},  and the squared unified-$(q, s)$ entropy \cite{Khan(2019)}.

More recently, a class of tight  $\alpha$-th power monogamy
relations were derived in multiqubit systems, such as the $\alpha$-th power of the concurrence ($C^{\alpha}$) \cite{Luo(2015)} with $\alpha\geq 2$ and $\alpha$-th power of EOF ($E^{\alpha}_f$) \cite{Fei1}  with $\alpha\geq \sqrt{2}$. Similar monogamy inequalities are discussed with the $\alpha$-th powers of the negativity \cite{Luo(2015)}, Tsallis-$q$ entropy \cite{Luo(2016)}, R\'{e}nyi-$q$ entropy \cite{Kim(2010)R}, and Unified-$(q, s)$ entropy entanglement \cite{KimBarry(2011)}. Other results are to tighten the monogamy inequalities which are useful for featuring multipartite entangled systems in entanglement distributions \cite{Fei3,Kim(2018)1,KimSC,Fei(2019)}.
 So far, although lots of monogamy relations are proposed \cite{T.J.Osborne,Kim(2009)N,Oliveira(2014),Bai3,Luo(2016),R(2015),Luo(2015),Bai(2014),Fei1,Fei3,Khan(2019),Kim(2018)1,KimSC,Fei(2019)}, there is a few unified monogamy relation for the aforementioned entanglement measures of bipartite systems \cite{Khan(2019)}. Our main goal in this paper is to present unified and tightened monogamy relations of entanglement measures encompassing the concurrence, the negativity, the entanglement of formation, Tsallis-$q$ entropy, and R\'{e}nyi-$q$ entropy, and Unified $(q,s)$ entropy for entangled qubit systems. The new monogamy inequalities with larger lower bounds also contain recent results \cite{Khan(2019),Fei3} as special cases. Moreover, we explore the generic feature of high-dimensional entangles systems going beyond qubit systems.

The outline of the rest is as follows. In Sec.II, we recall basic notations and definitions which will be used in the following sections. In Sec.III, we introduce the definition of the unified monogamy inequality. We prove a tightened monogamy inequality in Theorem 1 based on the unified monogamy inequality for tripartite entangled qubit systems. This result is then extended to general multi-partite entangled qubit systems in Theorem 2. A tighter inequality is further improved with special conditions on two-qubit entanglement as manifested in Theorem 3. We also derive several generalized monogamy inequalities based on the unified entanglement measures. In Sec.IV,  we prove a typical feature for high-dimension systems which violates the present monogamy relationship. Various examples for explaining our monogamy inequalities are included in Sec. 5 while the last section concludes the paper.

\section{Preliminaries}

In this section, we introduce the necessary notations and definitions which are useful throughout the paper. Let ${\cal H}_A$ and ${\cal H}_{B}$ be finite-dimensional Hilbert spaces. For any bipartite pure state $|\phi\rangle_{AB}$ or mixed state $\rho_{AB}$ on Hilbert space ${\cal H}_A\otimes {\cal H}_{B}$, there are lots of entanglement measures which will be explained as follows.

\subsection{Concurrence}

The concurrence of bipartite entanglement originates from the seminal work of Hill and Wootters \cite{Hill(1997),Wootters(1998)}. The concurrence is generated with the help of the superoperator that flips the spin of a qubit. Rungta et al. \cite{Rungta(2001)} generalized the spin-flip superoperator to a {\it universal inverter}, which acts on quantum systems of arbitrary dimension. The concurrence plays a major role in entanglement distribution protocols such as entanglement swapping and remote preparation of bipartite entangled states \cite{Gour(2004)}.

For an arbitrary bipartite pure state $|\phi\rangle_{AB}$ on Hilbert space ${\cal H}_A\otimes {\cal H}_{B}$, the concurrence \cite{Rungta(2001)} is given by
\begin{eqnarray}
C(|\phi\rangle_{AB})=\sqrt{2(1-{\rm{Tr}}(\rho^2_A))}
\label{eqn01}
\end{eqnarray}
where $\rho_A={\rm{Tr}}_B(|\phi\rangle_{AB}\langle\phi|)$ is the reduced density operator of the subsystem $A$ by tracing out the subsystem $B$.

For a mixed state $\rho_{AB}$, the concurrence is defined by the convex extension as
\begin{eqnarray}
C(\rho_{AB})=\inf_{\{p_i,|\phi_i\rangle\}}
\sum_ip_iC(|\phi_i\rangle_{AB})
\label{eq3}
\end{eqnarray}
where the infimum is over all possible pure state decompositions of $\rho_{AB}$, i.e. $\rho_{AB}=\sum_ip_i|\phi_i\rangle_{AB}\langle\phi_i|$, $\{p_i\}$ is a probability distribution with $p_i\geq 0$ and $\sum_ip_i=1$.

Remarkably, for a two-qubit mixed state $\rho$, the concurrence of $\rho$ can be evaluated as \cite{V. Coffman}:
\begin{eqnarray}
C(\rho)=\max\{0,\sqrt{\lambda_1}-\sqrt{\lambda_2}-\sqrt{\lambda_3}-\sqrt{\lambda_4}\}
\label{eqnC}
\end{eqnarray}
where $\lambda_i$s denote positive eigenvalues of the matrix $\rho(\sigma_y\otimes\sigma_y)\rho^*
(\sigma_y\otimes\sigma_y)$ with decreasing order,  $\sigma_y$ is Pauli matrix, and $\rho^*$ denotes the complex conjugate of $\rho$.

\subsection{Negativity}

The first criteria for the separability of quantum composed systems was given by Peres \cite{Peres(1996)}. He proved that the partial transposition of density operator $\rho^{T_A}_{AB}$ is positive if $\rho_{AB}$ is separable, this condition for the separability is called the positive partial transpose (PPT) criterion. Based on the trace norm of
the partial transpose $\rho^{T_A}_{AB}$ of the bipartite state $\rho_{AB}$, the negativity, a well-known quantification of bipartite entanglement is constructed.

For a bipartite state $\rho_{AB}$ on Hilbert space ${\cal H}_A\otimes {\cal H}_{B}$, its negativity \cite{Vidal(2002)} is defined as
\begin{eqnarray}
N(\rho_{AB})=\|\rho^{T_A}_{AB}\|_1-1
\label{eqn02}
\end{eqnarray}
where $\rho^{T_A}_{AB}$ is the partial transpose with respect to the subsystem $A$, and $\|X\|_1$ denotes the trace norm of $X$, i.e., $\|X\|_1=\rm{Tr}\sqrt{XX^+}$. If $\rho_{AB}$ is a bipartite pure state
in a $d\otimes d'(d\leq d')$ quantum system with the Schmidt decomposition,
$|\phi\rangle_{AB}=\sum^{d-1}_{i=0}\sqrt{\lambda_i}|ii\rangle$, $\lambda_i\geq0$, $\sum^{d-1}_{i=0}\lambda_i=1$.
$N(\rho_{AB})$ is given by $N(|\phi\rangle_{AB})=(\rm{Tr}\sqrt{\rho_{A}})^2-1$ \cite{Kim(2009)N}, where $\rho_A=Tr_B(|\phi\rangle_{AB}\langle\phi|)$. For a bipartite mixed state, the negativity is modified by the convex-roof extension, i.e., the convex-roof extended negativity (CREN), which is defined as
\begin{eqnarray}
N_c(\rho_{AB})=\inf_{\{p_i,|\phi_i\rangle\}}
\sum_{i}p_iN(|\phi_i\rangle_{AB})
\end{eqnarray}
where $\rho_{AB}=\sum_ip_i|\phi_i\rangle_{AB}\langle\phi_i|$,  $0\leq p_i\leq1$, and $\sum_ip_i=1$, the infimum is over all possible pure-state decompositions of $\rho_{AB}$. CREN gives a perfect discrimination between PPT bound entangled states and separable states in any bipartite quantum system \cite{Lee(2003)}.

Specially, any $n$-qubit pure state $|\psi\rangle_{A_1A_2\cdots A_n}$ on Hilbert space ${\cal H}_{A_1}\otimes{\cal H}_{A_2}\otimes\cdots\otimes {\cal H}_{A_n}$ can have a Schmidt decomposition with at most two nonzero Schmidt coefficients with respect to the bipartition of $A_1$ and all the others subsystems \cite{Kim(2009)N}. On the other hand, the negativity is equivalent to the concurrence for any pure states with Schmidt rank 2, i.e., $N(|\psi\rangle_{A_1|A_2\cdots A_n})=C(|\psi\rangle_{A_1|A_2\cdots A_n})$. Moreover, for any pure state $|\psi\rangle_{A_1A_2\cdots A_n}$, consider the reduced density operator $\rho_{A_iA_j}$ on two-qubit subsystems $A_i$, $A_j$. We have $N_c(\rho_{A_i|A_j})=C(\rho_{A_i|A_j})$ for any $i, j\in \{1, 2,\cdots, n\}, i \neq j$. It means that the negativity for any pure state with Schmidt rank 2, is equivalent to the entanglement measure of concurrence. Besides, there exist some certain relations between its concurrence and negativity \cite{Luo(2015),TT(2016)} for bipartite pure state. Hence, we only take use of the concurrence in what follows.

\subsection{The entanglement of formation}

For any bipartite pure system, Bennett et al. \cite{Bennett(1996)E} shown that the von Neumann entropy $S(\rho)=-\rm{Tr}(\rho\log_2\rho)$ of either subsystem is reasonable for featuring the entanglement of the composed system. This leads to the concept of entanglement of formation (EOF). EOF is independent of the subsystem being traced out, for a pure state $|\phi\rangle_{AB}$ on Hilbert space ${\cal H}_A\otimes {\cal H}_{B}$, EOF is defined as \cite{Bennett(1996)3824}:
\begin{eqnarray}
E_f(|\phi\rangle_{AB})=S(\rho_A)=-{\rm{Tr}}(\rho_A\log_2\rho_A)
\label{eqn03}
\end{eqnarray}
where $\rho_A={\rm Tr}_B(|\phi\rangle_{AB}\langle\phi|)$ denotes the density operator of the subsystem $A$ obtained by tracing out the subsystem $B$. EOF has operational meanings in entanglement preparation and data storage \cite{Bennett(1996)E}.

For a bipartite mixed state $\rho_{AB}$ on Hilbert space ${\cal H}_A\otimes {\cal H}_{B}$, EOF is given by
\begin{eqnarray}
E_f(\rho_{AB})=\inf_{\{p_i,|\phi_i\rangle\}}\sum_ip_iE(|\phi_i\rangle)
\end{eqnarray}
where the infimum takes over all possible pure-state decompositions of $\rho_{AB}=\sum_ip_i|\phi_i\rangle_{AB}\langle\phi_i|$.

Interestingly, EOF is related to the squared concurrence (SC) for an arbitrary quantum state of $2\otimes d$ systems \cite{Bai(2014)}. This relationship allows to evaluate the EOF of bipartite systems via its concurrence. Denote $f(x)=h(\frac{1+\sqrt{1-x}}{2})$, where $h(t)$ is binary entropy function given by $h(t)=-t\log_2(t)-(1-t)\log_2(1-t)$ for $0\leq t\leq 1$. For a pure state $|\phi\rangle_{AB}$ on $2\otimes d$ dimensional Hilbert space ${\cal H}_A\otimes {\cal H}_{B}$, EOF is given by \cite{Bai(2014)}:
\begin{eqnarray}
E_f(|\phi\rangle_{AB})=f(C^2(|\phi\rangle_{AB}))
\label{eqn04}
\end{eqnarray}
For two-qubit mixed state $\rho_{AB}$ we can get similar analytical formula.

\subsection{Tsallis-$q$ entropy}

As the one-parameter generalization of von Neumann entropy, Tsallis-$q$ entropy has been widely used in quantum information processing \cite{Tsallis(1988)}. It also provides useful conditions for the separability of quantum systems  \cite{Tsallis(2001)}, and can be used for characterizing classical statistical correlations inherent in quantum states \cite{Rajagopal(2005)}.

For a bipartite pure state $|\phi\rangle_{AB}$ on Hilbert space ${\cal H}_A\otimes {\cal H}_{B}$, Tsallis-$q$ entropy of an entanglement \cite{LV(1998),Kim(2010)T} is defined by
\begin{eqnarray}
T_q(|\phi\rangle_{AB})=T_q(\rho_A)=\frac{1}{q-1}(1-{\rm{Tr}}(\rho_A^q))
\label{eqn06}
\end{eqnarray}
where $\rho_A={\rm Tr}_B(|\phi\rangle_{AB}\langle\phi|)$ is the reduced density matrix by tracing over the subsystem $B$, $q>0$ and $q\neq 1$. When $q$ tends to 1, $T_q(\rho)$ converges to the von Neumann entropy, i.e., $\lim_{q \rightarrow1}T_q(\rho)=-{\rm Tr}\rho\log_2(\rho)=S(\rho)$ for any bipartite operator density $\rho$.

For a bipartite mixed state $\rho_{AB}$, Tsallis-$q$ entropy \cite{LV(1998)} is defined via the convex-roof extension as
\begin{eqnarray}
T_q(\rho_{AB})=\inf_{\{p_i,|\phi_i\rangle\}}\sum_ip_iT_q(|\phi_i\rangle_{AB})
\label{eqn07}
\end{eqnarray}
where the infimum is taken over all possible pure-state decompositions of $\rho_{AB}=\sum_ip_i|\phi_i\rangle_{AB}\langle\phi_i|$. One analytic relation between Tsallis-$q$ entropy and the concurrence is given by \cite{Yuan(2016)T}:
\begin{eqnarray}
T_q(|\phi\rangle_{AB})=f_q(C^2(|\phi\rangle_{AB}))
\label{eqn08}
\end{eqnarray}
for any $q$ satisfying $\frac{5-\sqrt{13}}{2}\leq q\leq\frac{5+\sqrt{13}}{2}$, where $|\phi\rangle_{AB}$ is a pure state of $2\otimes d(d\geq2)$ dimensional quantum system, and  $f_q(x)=\frac{1}{q-1}(1-(\frac{1+\sqrt{1-x}}{2})^q-(\frac{1-\sqrt{1-x}}{2})^q)$ is a monotonically increasing and convex function of $x$ with $0\leq x\leq 1$. A similar equality holds for any two-qubit mixed state $\rho_{AB}$ on Hilbert space ${\cal H}_A\otimes {\cal H}_{B}$.

\subsection{R\'{e}nyi-$q$ entropy}

For a quantum state $\rho$, its quantum R\'{e}nyi-$q$ entropy is given by
\begin{eqnarray}
S_q(\rho)=\frac{1}{1-q}\log_2 {\rm{Tr}} (\rho^q)
\label{}
\end{eqnarray}
for any $q>0$ with $q\neq 1$ \cite{HHH(1996)}. If $\rho$ has the spectral decomposition of
$\rho=\sum_i\lambda_i|\psi_i\rangle\langle\psi_i|$, we get $S_q(\rho)=H_q(X)$, where $X$ denotes the distribution probability of $X=\{\lambda_i\}$. We have that $\lim_{q \rightarrow1}S_q(\rho)=-{\rm  Tr}\rho\log_2(\rho)=S(\rho)$, thus, R\'{e}nyi-$q$ entropy is a generalization of von Neumann entropy.

As a generalization of EOF \cite{HHH(1996)}, the entanglement measure is defined by using the R\'{e}nyi-$q$ entropy of a bipartite pure state $|\phi\rangle_{AB}$ as
\begin{eqnarray}
R_q(|\phi\rangle_{AB})=S_q(\rho_A)=\frac{1}{1-q}\log_2 {\rm{Tr}} (\rho^q_A)
\label{eqn010}
\end{eqnarray}
Similar to the convex roof in Eq.(\ref{eqn07}), one entanglement measure is defined for a bipartite mixed state $\rho_{AB}$ as
\begin{eqnarray}
R_q(\rho_{AB})=\inf_{\{p_i,|\phi_i\rangle\}}\sum_ip_iR_q(|\phi_i\rangle_{AB})
\label{eqn011}
\end{eqnarray}
where the infimum is taken over all decompositions of $\rho_{AB}$ with pure states. In particular, for any two-qubit pure state $|\phi\rangle_{AB}$ with $q>0$, there exists an analytic formula of the entanglement measure with R\'{e}nyi-$q$ entropy \cite{Kim(2010)R} which is given by
\begin{eqnarray}
R_q(|\phi\rangle_{AB})=g_q(C^2(|\phi\rangle_{AB}))
\label{eqnrc1}
\end{eqnarray}
where $g_q(x)$ is defined by
$g_q(x)=\frac{1}{1-q}\log_2((\frac{1+\sqrt{1-x}}{2})^q+(\frac{1-\sqrt{1-x}}{2})^q)
$ with $0\leq x\leq 1$.
For a two-qubit mixed state $\rho_{AB}$, there exists a similar expression \cite{R(2015),R(2016)song} as Eq.(\ref{eqnrc1}).

\subsection{Unified-$(q, s)$ entropy}

Different from all the stated entropies, there exists a generalized entropy, i.e., Unified $(q, s)$-entropy \cite{Rathie(1991)}, which involves two real parameters $q$ and $s$:
\begin{eqnarray}
S_{q,s}(\rho)=\frac{1}{(1-q)s}({\rm{Tr}}(\rho^q)^s-1)
\label{}
\end{eqnarray}
where $q, s\geq0$ and $q\neq1$, $s\neq0$.  This two-parametric function includes R\'{e}nyi-$q$ entropy and Tsallis-$q$ entropy as the limiting case of $s\rightarrow0$ and $s\rightarrow1$, respectively. Moreover, for $q\to 1$, it converges to the von Neumann entropy \cite{KimBarry(2011)}.

Using the unified $(q, s)$-entropy with real parameters $q$ and $s$, for a bipartite pure state $|\phi\rangle_{AB}$ on Hilbert space ${\cal H}_A\otimes {\cal H}_{B}$, a bipartite entanglement measure is defined by \cite{KimBarry(2011)}:
\begin{eqnarray}
U_{q,s}(|\phi\rangle_{AB})=S_{q,s}(\rho_A)
\label{eqnUP}
\end{eqnarray}
for each $q, s\geq0$, where $\rho_A={\rm Tr}_B(|\phi\rangle_{AB}\langle\phi|)$ is the reduced density matrix of $|\phi\rangle_{AB}$ onto subsystem $A$. For a bipartite mixed state $\rho_{AB}$, its entanglement measure is given by
\begin{eqnarray}
U_{q,s}(\rho_{AB})=\inf_{\{p_i,|\phi_i\rangle\}}\sum_ip_iU_{q,s}(|\phi_i\rangle_{AB})
\label{eqnUH}
\end{eqnarray}
where the infimum is taken over all possible pure state decompositions of $\rho_{AB}=\sum_ip_i|\phi_i\rangle_{AB}\langle\phi_i|$ with  $\sum_ip_i=1$, and $p_i\geq0$.
It's worth pointing out that there exists a functional relation \cite{Khan(2019)}
\begin{eqnarray}
U_{q,s}(|\phi\rangle_{AB})=f_{q,s}(C^2(|\phi\rangle_{AB}))
\label{eqnfqs1}
\end{eqnarray}
for any $2\otimes d$ dimensional pure state $|\phi\rangle_{AB}$,
where $f_{q,s}(x)=((1+\sqrt{1-x})^q+(1-\sqrt{1-x})^q)^s-2^{qs})/(s(1-q)2^{qs})$ with $0\leq x \leq1$ and $(q, s)\in \mathbb{\textit{R}}$. Similar relation holds for two-qubit mixed states.

\section{Monogamy relation of multipartite entangled qubit systems}

Denote $\rho_{AB_0\cdots B_{n-1}}$ as the state of a multipartite system on finite dimensional Hilbert space ${\cal H}_A\otimes {\cal H}_{B_0}\otimes\cdots \otimes {\cal H}_{B_{n-1}}$, where ${\cal H}_A$ and ${\cal H}_{B_i}$ are finite dimensional Hilbert spaces.

{\it Definition 1}. Given a bipartite entanglement measure $E$ of the quantum states $\rho_{AB_0\cdots B_{n-1}}$, $E^{\alpha_{c}}$ is said to be monogamous if the following inequality holds
\begin{eqnarray}
E^{\alpha_{c}}(\rho_{A|B_0\cdots B_{n-1}})\geq \sum_{i=0}^{n-1} E^{\alpha_{c}}(\rho_{A|B_{i}})
\label{eqn1}
\end{eqnarray}
where $E(\rho_{A|B_0\cdots B_{n-1}})$ is the entanglement measure of $\rho_{A|B_0\cdots B_{n-1}}$ which is reduced density matrix with respect to the bipartition $A$ and $B_0\cdots B_{n-1}$, and $E(\rho_{A|B_i})$ is the entanglement measure of the reduced density matrix $\rho_{A|B_i}={\rm Tr}_{B_0\cdots B_{i-1}B_{i+1}\cdots B_{n-1}}(\rho_{AB_0\cdots B_{n-1}})$ of subsytems $A, B_i$ for $i=0, 1, 2, \cdots, n-1$, $\alpha_{c}$ is the minimum positive number for $E^{\alpha_{c}}$ to be monogamous \cite{GYG19}.

Generally, for an entanglement measure $E$ one can get a quantity $E^{\alpha_{c}}$ satisfying the monogamy inequality (\ref{eqn1}) for entangled qubit systems even if $E$ is not monogamous. Different values of $\alpha_{c}$ exist for aforementioned entanglement measures \cite{T.J.Osborne,Kim(2009)N,Oliveira(2014),Bai3,Bai(2014),Luo(2016),R(2015),Khan(2019)}. Similar result cannot be proved for high-dimensional systems. Hence, in this section, we consider all qubit systems while the different result will be proved in the section for high-dimensional systems.

\subsection{Tripartite entangled qubit systems}

In this subsection, we present the first result which states a class of tight monogamy inequalities for tripartite entangled qubit systems.

\textbf{Theorem 1}.  For an arbitrary three-qubit state $\rho_{A|B_0B_1}$ on Hilbert space ${\cal H}_A\otimes {\cal H}_{B_0}\otimes {\cal H}_{B_1}$, assume that $E^{\alpha_{c}}$ is a monogamous entanglement measure, we obtain that
\begin{eqnarray}
E^\alpha(\rho_{A|B_0B_1})&\geq& ((\gamma_0+1)^t-\gamma_0^t)E^\alpha(\rho_{A|B_1})
\nonumber\\
&&+E^\alpha(\rho_{A|B_0})
 \label{eqn6}
\end{eqnarray}
for $\alpha\geq \alpha_{c}$, where $t=\alpha/{\alpha_{c}}$, $\gamma_0=\gamma^{\alpha_{c}}$, and $\gamma$ is a constant satisfying that $E(\rho_{A|B_0})\geq \gamma E(\rho_{A|B_1})$ for $\gamma\geq1$.

{\bf Proof}. For a tripartite entangled qubit state  $\rho_{A|B_0B_1}$ on Hilbert space ${\cal H}_A\otimes {\cal H}_{B_0}\otimes {\cal H}_{B_1}$ with reduced density operators $\rho_{A|B_0}$ on Hilbert space ${\cal H}_A\otimes {\cal H}_{B_0}$ and $\rho_{A|B_1}$ on Hilbert space ${\cal H}_A\otimes {\cal H}_{B_1}$, from the inequality (\ref{eqn1}) it follows that
\begin{eqnarray}
E^\alpha(\rho_{A|B_0B_1})
&\geq& [E^{\alpha_{c}}(\rho_{A|B_0})+E^{\alpha_{c}}(\rho_{A|B_1})]^t
\nonumber
\\
&=&E^\alpha(\rho_{A|B_0})(1+\frac{E^{\alpha_{c}}(\rho_{A|B_1})}{E^{\alpha_{c}}(\rho_{A|B_0})})^t
\label{eqn8}
\end{eqnarray}
Note that
\begin{eqnarray}
 (1+\frac{E^{\alpha_{c}}(\rho_{A|B_1})}{E^{\alpha_{c}}(\rho_{A|B_0})})^t
 &\geq &
 1+((\gamma_0+1)^t-\gamma_0^t)
\nonumber \\
&&\times (\frac{E^{\alpha_{c}}(\rho_{A|B_1})}{E^{\alpha_{c}}
(\rho_{A|B_0})})^t
 \label{eqn9}
\end{eqnarray}
when $E(\rho_{A|B_0})\geq \gamma E(\rho_{A|B_1})$. The inequality (\ref{eqn9}) is due to the inequality: $(1+x)^t\geq1+((\gamma_0+1)^t-\gamma_0^t)x^t$ for any $0\leq x\leq \frac{1}{\gamma_0}, \gamma_0\geq1$, and $t\geq1$. In fact, this inequality can be easily proved as follows: If $x=0$, the inequality is trivial. Otherwise, let $f(t,x)=\frac{(1+x)^t-1}{x^t}$, with $t\geq 1$, $\gamma_0\geq1$ and $0\leq x\leq \frac{1}{\gamma_0}$. From $\frac{\partial f}{\partial x}=\frac{tx^{t-1}(1-(1+x)^{t-1})}{x^{2t}}$, it is easy to check that $(1+x)^{t-1}\geq1$. Thus $\frac{\partial f}{\partial x}\leq0$, which implies that $f(t,x)$ is a decreasing function of $x$, and $f(t,x)\geq f(t,\frac{1}{\gamma_0})=(\gamma_0+1)^t-\gamma_0^t$. Hence, the inequalities (\ref{eqn8}) and (\ref{eqn9}) lead to
\begin{eqnarray}
E^\alpha(\rho_{A|B_0B_1})
&\geq & E^\alpha(\rho_{A|B_0})(1+((\gamma_0+1)^t-\gamma_0^t)
\nonumber\\
&&
\times\frac{(E^{\alpha_{c}}
  (\rho_{A|B_1}))^t}{(E^{\alpha_{c}}(\rho_{A|B_0}))^t})
\nonumber\\
&=&E^\alpha(\rho_{A|B_0})+((\gamma_0+1)^t-\gamma_0^t)
\nonumber\\
&&\times{}E^\alpha(\rho_{A|B_1})
 \label{eqn11}
\end{eqnarray}
which has completed the proof. $\hfill\square$

\subsection{Multipartite entangled qubit systems}

A natural question of Theorem 1 is whether it holds for multi-qubit states. Actually, we prove a generalized relationship for multi-partite entangled systems. Our main result of the monogamy relation in the following section is based on the Hamming weight of binary representations of proper vectors.

{\it Definition 2}. For any non-negative integer $j$ and its binary expansion is given by $j=\sum_{s=0}^{m-1}j_s2^s$
with $\log_2j\leq m$ and  $j_s\in\{0,1\}$ for $s=0, 1, \cdots, m-1$. We can define a unique binary vector $\vec{j}$ associated with $j$ as $\vec{j}=(j_0,j_1,\cdots, j_{m-1})$.
Its \textit{Hamming weight}, $\omega_H(\vec{j})$, is defined as the number of 1's, i.e., the number of 1's in $\{j_0,j_1,\cdots, j_{m-1}\}$ \cite{Kim(2018)1,KimSC,Fei(2019)}.

\textbf{Theorem 2.} Consider an arbitrary $n+1$-qubit state $\rho_{A|{\bf B}}$ with ${\bf B}=B_0\cdots{}B_{n-1}$ on Hilbert space ${\cal H}_{A}\otimes {\cal H}_{B_0}\otimes \cdots \otimes {\cal H}_{B_{n-1}}$. For an entanglement measure $E$, assume that $E^{\alpha_{c}}$ is a monogamous entanglement measure with some constant $\alpha_c$. Then, the entanglement measure satisfies
\begin{eqnarray}
E^\alpha(\rho_{A|{\bf B}})\geq\sum_{j=0}^{n-1} ((\gamma_0+1)^t-\gamma_0^t)^{\omega_H(\vec{j})} E^\alpha(\rho_{A|B_j})
\label{eqn14}
\end{eqnarray}
for $\alpha\geq \alpha_{c}$, where $t=\alpha/{\alpha_{c}}$, $\gamma_0=\gamma^{\alpha_{c}}$, and $\gamma(\geq1)$ is a constant satisfying the inequality $E(\rho_{A|B_j})\geq \gamma E(\rho_{A|B_{j+1}})$ for $j=0, 1, \cdots, n-2$.

{\bf Proof}. For Eq.(\ref{eqn1}), without loss of generality, for each quantum entanglement measure $E$ satisfying the following inequality
\begin{eqnarray}
 E(\rho_{A|B_{j}})\geq  E(\rho_{A|B_{j+1}})\geq0
 \label{eqn5}
\end{eqnarray}
by reordering and relabeling all subsystems $A, B_i$s, where $\rho_{A|B_0\cdots{}B_{n-1}}$ is a joint system  on Hilbert space ${\cal H}_{A}\otimes {\cal H}_{B_0}\otimes \cdots{}\otimes {\cal H}_{B_{n-1}}$, and $\rho_{A|B_{j}}$  are reduced density operators of $\rho_{A|B_0\cdots{}B_{n-1}}$ on Hilbert space ${\cal H}_{A}\otimes {\cal H}_{B_j}$. For example, we can define
\begin{eqnarray}
 &&\!\!\!\!\!\!\!\!\!E(\rho_{A|B_{i_0}})=\max\{E(\rho_{A|B_i}), \forall i=0, \cdots, n-1\}
 \nonumber
 \\
&&\!\!\!\!\!\!\!\!\!E(\rho_{A|B_{i_{k}}})=\max\{E(\rho_{A|B_i}),\forall i,i\neq i_0, \cdots, i_{k-1}\}
\end{eqnarray}
with $ k\leq n-1$. Henceforth, $E(\rho_{A|B_{j}})$ satisfy the inequality (\ref{eqn1}) as
\begin{eqnarray}
E^{\alpha_{c}}(\rho_{A|B_0\cdots B_{n-1}})\geq \sum_{j=0}^{n-1} E^{\alpha_{c}}(\rho_{A|B_{j}})
\label{eqn55}
\end{eqnarray}
where $E^{\alpha_{c}}$ is monogamous entanglement measure for some constant $\alpha_c$. Due to the monotonicity of the function: $f(x)=x^\alpha$ for $\alpha\geq 1$, we get
\begin{eqnarray}
E^{\alpha}(\rho_{A|{\bf B}})\geq (\sum_{j=0}^{n-1}E^{\alpha_{c}}(\rho_{A|B_j}))^t
\label{eqn15}
\end{eqnarray}
from the inequality (\ref{eqn55}). Hence, it is sufficient to show that
\begin{eqnarray}
(\sum_{j=0}^{n-1}E^{\alpha_{c}}(\rho_{A|B_j}))^t
&\geq& \sum_{j=0}^{n-1} ((\gamma_0+1)^t-\gamma_0^t)^{\omega_H(\vec{j})}
\nonumber
\\
&&\times{} E^\alpha(\rho_{A|B_j})
\label{eqn16}
\end{eqnarray}

We first prove the inequality (\ref{eqn14}) for the case of $n=2^k$ by induction on $n$. And then, we extend the result for any positive integer $n$. Actually, from Theorem 1 the inequality (\ref{eqn14}) holds for $k=1$, i.e., $n=2$. Assume the inequality (\ref{eqn14}) is true for $n=2^{k-1}$ with $k\geq 2$. Consider the case of $n=2^k$. For a $(n+1)$-partite state $\rho_{A|{\bf B}}$ on Hilbert space ${\cal H}_{A}\otimes {\cal H}_{B_0}\otimes \cdots \otimes {\cal H}_{B_{n-1}}$ and its reduced density matrices $\rho_{A|B_j}$ on Hilbert space ${\cal H}_{A}\otimes {\cal H}_{B_j}$ with $j=0, 1, \cdots, n-1$, we have
\begin{eqnarray}
(\sum_{j=0}^{n-1}E^{\alpha_{c}}(\rho_{A|B_j}))^t
 &=& (1+\frac{\sum_{j=2^{k-1}}^{2^k-1}E^{\alpha_{c}}(\rho_{A|B_j})}{\sum_{j=0}^{2^{k-1}-1}E^{\alpha_{c}}(\rho_{A|B_j})})^t
\nonumber\\
&& \times(\sum_{j=0}^{2^{k-1}-1}E^{\alpha_{c}}(\rho_{A|B_j}))^t
\label{eqn17}
\end{eqnarray}

Due to the inequality $E(\rho_{A|B_j})\geq \gamma E(\rho_{A|B_{j+1}})$, it follows that
\begin{eqnarray}
0\leq\frac{\sum_{j=2^{k-1}}^{2^k-1}
E^{\alpha_{c}}
(\rho_{A|B_j})}{\sum_{j=0}^{2^{k-1}-1}
E^{\alpha_{c}}(\rho_{A|B_j})}\leq\frac{1}{\gamma_0}
\label{eqn18}
\end{eqnarray}
Thus, Eqs.(\ref{eqn17}) and (\ref{eqn18}) yield to
\begin{eqnarray}
\!\!\!\!\!\!\!\!\!\Delta&:=&(\sum_{j=0}^{n-1}E^{\alpha_{c}}(\rho_{A|B_j}))^t
\nonumber\\
&=&(1+\frac{\sum_{j=2^{k-1}}^{2^k-1}E^{\alpha_{c}}
(\rho_{A|B_j})}{\sum_{j=0}^{2^{k-1}-1}
E^{\alpha_{c}}(\rho_{A|B_j})})^t
(\sum_{j=0}^{2^{k-1}-1}E^{\alpha_{c}}(\rho_{A|B_j}))^t
 \nonumber\\
&\geq&(\sum_{j=0}^{2^{k-1}-1}E^{\alpha_{c}}
(\rho_{A|B_j}))^t(1+((\gamma_0+1)^t-\gamma_0^t)
 \nonumber\\
&& \times(\frac{\sum_{j=2^{k-1}}^{2^k-1}
E^{\alpha_{c}}(\rho_{A|B_j})}{
\sum_{j=0}^{2^{k-1}-1}E^{\alpha_{c}}(\rho_{A|B_j})})^t)
 \nonumber\\
&=&(\sum_{j=0}^{2^{k-1}-1}E^{\alpha_{c}}(\rho_{A|B_j}))^t
+((\gamma_0+1)^t-\gamma_0^t)
 \nonumber\\
&&
\times{} (\sum_{j=2^{k-1}}^{2^k-1}
E^{\alpha_{c}}(\rho_{A|B_j}))^t
\label{eqn19}
\end{eqnarray}

From the induction hypothesis, it follows that
\begin{eqnarray}
(\sum_{j=0}^{2^{k-1}-1}E^{\alpha_{c}}(\rho_{A|B_j}))^t
&\geq& \sum_{j=0}^{2^{k-1}-1}
((\gamma_0+1)^t-\gamma_0^t)^{\omega_H(\vec{j})}
\nonumber
\\
&&\times{}E^\alpha(\rho_{A|B_j})
\label{eqn20}
\end{eqnarray}
Moreover, the last summation in the inequality (\ref{eqn19}) is a summation of $2^{k-1}$ terms starting from $j=2^{k-1}$ to $j=2^k-1$. Thus, after possible indexing and reindexing subsystems, the induction hypothesis leads us to
\begin{eqnarray}
(\sum_{j=2^{k-1}}^{2^k-1}E^{\alpha_{c}}(\rho_{A|B_j}))^t
&\geq& \sum_{j=2^{k-1}}^{2^k-1}
((\gamma_0+1)^t-\gamma_0^t)^{\omega_H(\vec{j})-1}
\nonumber\\
&&\times{}E^\alpha(\rho_{A|B_j})
\label{eqn21}
\end{eqnarray}

From the inequalities (\ref{eqn19})-(\ref{eqn21}), we have
\begin{eqnarray}
 (\sum_{j=0}^{2^k-1}E^{\alpha_{c}}(\rho_{A|B_{j}}))^t
 &\geq&\sum_{j=0}^{2^k-1} ((\gamma_0+1)^t-\gamma_0^t)^{\omega_H(\vec{j})}
\nonumber\\
 &&\times E^\alpha(\rho_{A|B_j})
\end{eqnarray}
which recovers the inequality (\ref{eqn14}).

Now,  we show that an $(n+1)$-qubit state satisfies monogamy inequality (\ref{eqn14}) for arbitrary positive integer $n$.  Note that one can always consider a power of 2 be an upper bound of $n$, that is, $0\leq n \leq 2^k$ for some integer $k$. We also consider a $(2^k+1)$-partite quantum state
$\varrho_{AB_0B_1 \cdots B_{2^k-1}}$ on Hilbert space ${\cal H}_{A}\otimes {\cal H}_{B_0}\otimes \cdots \otimes {\cal H}_{B_{2^k-1}}$ defined by
\begin{eqnarray}
\varrho_{AB_0B_1 \cdots B_{2^k-1}}=\rho_{A{\bf B}}\otimes\rho_{B_n \cdots B_{2^k-1}}
\label{eqn23}
\end{eqnarray}
where $\rho_{A{\bf B}}$ is density operator of subsystems $A, B_0, \cdots, B_{n-1}$ on Hilbert space ${\cal H}_{A}\otimes {\cal H}_{B_0}\otimes \cdots \otimes {\cal H}_{B_{n-1}}$, and $\rho_{B_n \cdots B_{2^k-1}}$ is density operator of subsystems $B_n, \cdots, B_{2^k-1}$ on Hilbert space ${\cal H}_{B_n}\otimes \cdots \otimes {\cal H}_{B_{2^k-1}}$. Since $\varrho_{AB_0B_1 \cdots B_{2^k-1}}$ is a $(2^k+1)$-partite quantum state, as its proved above we have that
\begin{eqnarray}
E^\alpha(\varrho_{A|B_0B_1 \cdots B_{2^k-1}})
&\geq& \sum_{j=0}^{2^k-1} ((\gamma_0+1)^t-\gamma_0^t)^{\omega_H(\vec{j})}
\nonumber
\\
&&\times{}E^\alpha(\varrho_{A|B_j})
\label{eqn24}
\end{eqnarray}
where $\varrho_{A|B_j}$ are reduced density operators of $\varrho_{A|B_0B_1 \cdots B_{2^k-1}}$, $j=0, 1, \cdots, 2^k-1$.

Note that $\varrho_{A|B_0B_1 \cdots B_{2^k-1}}$ is separable. We get $E(\varrho_{A|B_0B_1 \cdots B_{2^k-1}})= E(\rho_{A|{\bf B}})$ which implies that $E(\varrho_{A|B_j})=0$, for $j=n, \cdots, 2^k-1$. It follows that $\varrho_{AB_j}=\rho_{AB_j}$, for $j=0, \cdots, n-1$. This leads us to
\begin{eqnarray}
E^\alpha(\rho_{A|{\bf B}})&=&E^\alpha(\varrho_{A|B_0B_1 \cdots B_{2^k-1}})
\nonumber\\
&\geq &\sum_{j=0}^{2^k-1} ((\gamma_0+1)^t-\gamma_0^t)^{\omega_H(\vec{j})} E^\alpha(\varrho_{A|B_j})
\nonumber\\
&=&\sum_{j=0}^{n-1} ((\gamma_0+1)^t-\gamma_0^t)^{\omega_H(\vec{j})}
 \nonumber\\
 &&\times{}E^\alpha(\rho_{A|B_j})
\label{eqn25}
\end{eqnarray}
This completes the proof. $\hfill\square$

When $E(\rho_{A|B_j})\geq \gamma E(\rho_{A|B_{j+1}})$, i.e., $\gamma_0=\gamma^{\alpha_{c}}=1$, the reduced inequality is tighter than recent result \cite{Khan(2019)} as a special case. The monogamy inequality in Theorem 2 can be further improved with special conditions on bipartite entanglement measures.

\textbf{Theorem 3.} For an $n+1$-qubit state $\rho_{A|{\bf B}}$ with ${\bf B}=B_0\cdots{}B_{n-1}$ on Hilbert space ${\cal H}_{A}\otimes {\cal H}_{B_0}\otimes \cdots \otimes {\cal H}_{B_{n-1}}$ and a monogamous entanglement measure $E^{\alpha_{c}}$ in terms of entanglement measure $E$, we have
\begin{eqnarray}
E^\alpha(\rho_{A|{\bf B}})\geq\sum_{j=0}^{n-1} ((\gamma_0+1)^t-\gamma_0^t)^{j} E^\alpha(\rho_{A|B_j})
\label{eqn26}
\end{eqnarray}
where $E$ satisfies the inequality
$E(\rho_{A|B_{i}})\geq \gamma\sum_{j=i+1}^{n-1} E(\rho_{A|B_{j}})$
 for $i=0, 1, \cdots, n-2$, $t=\alpha/{\alpha_{c}}$, $\alpha\geq \alpha_{c}$, and $\gamma_0=\gamma^{\alpha_{c}}$ with $\gamma\geq1$.

{\bf Proof}. It is enough to prove
\begin{eqnarray}
 (\sum_{j=0}^{n-1}E^{\alpha_{c}}(\rho_{A|B_j}))^t\geq \sum_{j=0}^{n-1} ((\gamma_0+1)^t-\gamma_0^t)^{j} E^\alpha(\rho_{A|B_j})
\label{eqn27}
\end{eqnarray}
where $\rho_{A|{\bf B}}$ is a multipartite state on Hilbert space ${\cal H}_{A}\otimes {\cal H}_{B_0}\otimes \cdots \otimes {\cal H}_{B_{n-1}}$ with two-qubit reduced density matrices $\rho_{A|B_{j}}$, $j=0, 1, \cdots, n-1$. This can be proved by induction on $n$.

Note that the inequality (\ref{eqn11}) guarantees the validity of the inequality (\ref{eqn27}) for $n=2$. Assume that the inequality (\ref{eqn27}) is valid for any positive integer less than $n$. Consider a  multipartite quantum state $\rho_{A|{\bf B}}$ on Hilbert space ${\cal H}_{A}\otimes {\cal H}_{B_0}\otimes \cdots \otimes {\cal H}_{B_{n-1}}$ and its two-qubit reduced density operators $\rho_{A|B_j}$ on Hilbert space ${\cal H}_{A}\otimes {\cal H}_{B_j}$. It is easy to verify that
\begin{eqnarray}
0\leq\frac{\sum\limits_{j=i+1}^{n-1} E^{\alpha_{c}}(\rho_{A|B_{j}})}{ E^{\alpha_{c}}(\rho_{A|B_{i}})}\leq \frac{1}{\gamma_0}
\label{eq28}
\end{eqnarray}
Thus, using the inequality: $(1+x)^t\geq1+((\gamma_0+1)^t-\gamma_0^t)x^t$, we find that
\begin{eqnarray}
\Delta&:=&(\sum_{j=0}^{n-1} E^{\alpha_{c}}(\rho_{A|B_{j}}))^t
\nonumber\\
&=&E^{\alpha}(\rho_{A|B_0})(1+\frac{\sum_{j=1}^{n-1} E^{\alpha_{c}}(\rho_{A|B_j})}{
E^{\alpha_{c}}(\rho_{A|B_0})})^t
\nonumber\\
&\geq & E^{\alpha}(\rho_{A|B_0})(1+((\gamma_0+1)^t-\gamma_0^t)
\nonumber\\
&&\times{}(\frac{\sum_{j=1}^{n-1}E^{\alpha_{c}}
(\rho_{A|B_{j}})}{E^{\alpha_{c}}(\rho_{A|B_0})})^t)
\nonumber\\
&=&E^{\alpha}(\rho_{A|B_0})+((\gamma_0+1)^t-\gamma_0^t)
\nonumber\\
&&\times{}(\sum_{j=1}^{n-1} E^{\alpha_{c}}(\rho_{A|B_j}))^t
\label{eqn268}
\end{eqnarray}
Hence, the induction hypothesis leads us to
\begin{eqnarray}
 (\sum_{j=1}^{n-1} E^{\alpha_{c}}(\rho_{A|B_j}))^t&\geq& \sum_{j=1}^{n-1} ((\gamma_0+1)^t-\gamma_0^t)^{j-1} \nonumber\\
&&\times{}E^\alpha(\rho_{A|B_j})
 \label{eqn278}
\end{eqnarray}

Finally, the inequalities (\ref{eqn268}) and (\ref{eqn278}) recover the inequality (\ref{eqn27}). Combining the inequalities (\ref{eqn15}) with (\ref{eqn27}), we obtain the inequality (\ref{eqn26}). $\hfill\square$

When $ E(\rho_{A|B_{i}})\geq \sum_{j=i+1}^{n-1} E(\rho_{A|B_{j}})$, i.e., $\gamma_0=\gamma^{\alpha_{c}}=1$, the inequality (\ref{eqn26}) is reduced to the following inequality \cite{Fei3}
\begin{eqnarray}
E^{\alpha_{c}}(\rho_{A|B_0\cdots B_{n-1}})\geq
\sum_{i=0}^{n-1}(2^t-1)^iE^\alpha(\rho_{A|B_i})
\label{eqn12}
\end{eqnarray}

Generally, the inequality (\ref{eqn26}) in Theorem 3 is tighter than the inequality (\ref{eqn14}) in Theorem 2 for $\alpha\geq \alpha_{c}$ for any multipartite quantum state $\rho_{A|{\bf B}}$, the reasons are as follows. For any nonnegative integer $j$ and its binary vector $\vec{j}$, the Hamming weight $\omega_H(\vec{j})$ is bounded above by $\log_2j$. It follows that $\omega_H(\vec{j})\leq \log_2j\leq j$. Therefore, we get
 \begin{eqnarray}
E^\alpha(\rho_{A|{\bf B}})
&\geq& \sum_{j=0}^{n-1} ((\gamma_0+1)^t-\gamma_0^t)^{j}
   E^\alpha(\rho_{A|B_j})
\nonumber\\
& \geq&\sum_{j=0}^{n-1} ((\gamma_0+1)^t-\gamma_0^t)^{\omega_H(\vec{j})} E^\alpha(\rho_{A|B_j})
\end{eqnarray}

\subsection{Multipartite qubit systems with generalized bipartitions}

In Subsec.C, we get the monogamy relation for a joint state with the bipartite partition $A$ and $B_0\cdots B_{n-1}$. Now, we derive a generalized monogamy inequality for unifying entanglement measure $E(\rho_{\bf {A|B}})$ on $m+n$-partite system under the partition ${\bf {A}}=A_1A_2\cdots A_m$ and ${\bf {B}}=B_1B_2\cdots B_n$.  With this unified monogamy relation, we present two generalized inequalities. Additionally, we will prove an upper bound of $C^\alpha(|\psi\rangle_{\bf {A|B}})$$(0\leq\alpha\leq2)$ and $U_{q,s}^\alpha(|\psi\rangle_{\bf {A|B}})$$(0\leq\alpha\leq1)$. Here, $\rho_{\bf {A|B}}, |\psi\rangle_{\bf {A|B}}$ is defined by $\rho_{\bf {A|B}}=\rho_{A_1\cdots A_mB_1\cdots B_n}$ and $|\psi\rangle_{\bf {A|B}}=|\psi\rangle_{A_1\cdots A_mB_1\cdots B_n}$, respectively, $\rho_{{\bf A}}$ are on Hilbert space ${\cal H}_{A_1}\otimes {\cal H}_{A_2}\otimes \cdots \otimes {\cal H}_{A_{m}}$ and $\rho_{{\bf B}}$ are on Hilbert space ${\cal H}_{B_1}\otimes {\cal H}_{B_2}\otimes \cdots \otimes {\cal H}_{B_{n}}$.

According to the inequality (\ref{eqn55}), it is straightforward to prove the following Theorems.

\textbf{Theorem 4}. For any $m+n$-partite qubit state $\rho_{\bf {A|B}}$ on Hilbert space ${\cal H}_{A_1}\otimes \cdots \otimes {\cal H}_{A_m}\otimes{\cal H}_{B_1}\otimes \cdots \otimes {\cal H}_{B_{n}}$, each monogamous entanglement measure $E^{\alpha_{c}}$ satisfies
\begin{eqnarray}
E^{\alpha_c}(\rho_{\bf {A|B}})\geq \sum^m_{i=1}\sum^n_{j=1}E^{\alpha_c}(\rho_{A_i|B_j})
\label{eqntg}
\end{eqnarray}
where $\rho_{A_i|B_j}$ are reduced density operators of $\rho_{\bf {A|B}}$ on Hilbert space ${\cal H}_{A_i}\otimes {\cal H}_{B_j}$, $i=1, \cdots, m; j=1, \cdots, n$.

\textbf{Theorem 5}. For any $m+n$-qubit state $\rho_{\bf {A|B}}$ on Hilbert space ${\cal H}_{A_1}\otimes \cdots \otimes {\cal H}_{A_m}\otimes{\cal H}_{B_1}\otimes \cdots \otimes {\cal H}_{B_{n}}$, each monogamous entanglement measure $E^{\alpha_{c}}$ satisfies
\begin{eqnarray}
E^{\alpha}(\rho_{\bf {A|B}})\geq \sum^m_{i=1}\sum^n_{j=1}E^{\alpha}(\rho_{A_i|B_j})
\label{Theorem 5}
\end{eqnarray}
for $\alpha\geq \alpha_c$, where $\rho_{A_i|B_j}$ are reduced density operators of $\rho_{\bf {A|B}}$ on Hilbert space ${\cal H}_{A_i}\otimes {\cal H}_{B_j}$, $i=1, \cdots, m; j=1, \cdots, n$.

In order to get a tightened monogamy inequality than the inequality (\ref{Theorem 5}) in Theorem 5, we present the following Lemma 1 firstly.

\textbf{Lemma 1}. For $a_1\geq a_2\geq\cdots a_n\geq0$, and $\mu\geq1$, then the following inequality holds
\begin{eqnarray}
(\sum_{i=1}^na_i)^\mu&\geq& \sum_{i=0}^{n-1}((i+1)^\mu-i^\mu)a_{i+1}^\mu
\label{eqnlemma1}
\end{eqnarray}

The proof of Lemma 1 is firstly presented in ref.\cite{GYG19}. For each entanglement measure $E$, by relabeling the subsystems we get $E(\rho_{A_iB_j})\geq E(\rho_{A_iB_{j+1}})\geq E(\rho_{A_{i+1}B_{1}})$ for $i=1, 2, \cdots,  m-1$, $j=1,2\cdots n-1$. Combined with Lemma 1, we obtain the following Theorem with a larger lower bound than the existing monogamy relations in Theorem 5.

\textbf{Theorem 6}. For any $m+n$-qubit state $\rho_{\bf {A|B}}$ on Hilbert space ${\cal H}_{A_1}\otimes \cdots \otimes {\cal H}_{A_m}\otimes{\cal H}_{B_1}\otimes \cdots \otimes {\cal H}_{B_{n}}$, assume that the monogamous entanglement measure $E^{\alpha_{c}}$ satisfies the inequality (\ref{eqntg}). We get
\begin{eqnarray}
E^{\alpha}(\rho_{\bf {A|B}})&\geq&
\sum_{j=1}^{n}
\sum_{i=0}^{m-1}((ni+j)^t-(ni+j-1)^t)
\nonumber\\
&&\times E^{\alpha}(\rho_{A_{i+1}|B_j})
\end{eqnarray}
for $\alpha\geq \alpha_c$ and $t=\alpha/{\alpha_{c}}$, $\alpha\geq \alpha_{c}$, where $\rho_{A_i|B_j}$ are reduced density operators of $\rho_{\bf {A|B}}$ on Hilbert space ${\cal H}_{A_i}\otimes {\cal H}_{B_j}$, $i=1, \cdots, m; j=1, \cdots, n$.

{\bf Proof}. From the inequality (\ref{eqntg}) we get
\begin{eqnarray}
E^{\alpha}(\rho_{\bf {A|B}})\geq (\sum^m_{i=1}\sum^n_{j=1}E^{\alpha_c}(\rho_{A_iB_j}))^t
\label{eqntg1}
\end{eqnarray}
Using the inequalities of $E(\rho_{A_iB_j})\geq E(\rho_{A_iB_{j+1}})\geq E(\rho_{A_{i+1}B_{1}})$ for $i=1, 2, \cdots, m-1$; $j=1,2\cdots n-1$, from Lemma 1 it follows that
\begin{eqnarray}
E^{\alpha}(\rho_{\bf {A|B}})&\geq&
\sum^n_{j=1}(j^t-(j-1)^t)E^{\alpha}(\rho_{A_1|B_j})
\nonumber\\
&&+\sum^n_{j=1}((n+j)^t-(n+j-1)^t)E^{\alpha}(\rho_{A_2|B_j})
\nonumber\\
&&+\cdots+\sum^n_{j=1}((mn-n+j)^t
\nonumber\\
&&-(mn-n+j-1)^t)\times E^{\alpha}(\rho_{A_m|B_j})
\nonumber\\&=&
\sum_{j=1}^{n}
\sum_{i=0}^{m-1}((ni+j)^t-(ni+j-1)^t)
\nonumber\\
&&\times E^{\alpha}(\rho_{A_{i+1}|B_j})
\end{eqnarray}
which completes the proof.  $\hfill\square$

The concurrence \cite{Linear entropy,Linear entropy(2008)} is related to the linear entropy of a state $\rho$ as
\begin{eqnarray}
T(\rho)=1-{\rm tr}(\rho^2)
\label{eqnLinear1}
\end{eqnarray}
for any bipartite state $\rho$. Moreover, $T(\rho)$ has the property
\begin{eqnarray}
T(\rho_{AB})\leq T(\rho_{A})+T(\rho_{B})
\label{eqnLinear2}
\end{eqnarray}
From this relation we get following Theorem 7 about the concurrence related to the linear entropy.

\textbf{Theorem 7}. For any $m+n$-qubit pure state $|\psi\rangle_{\bf{AB}}$ on Hilbert space ${\cal H}_{A_1}\otimes \cdots \otimes {\cal H}_{A_m}\otimes{\cal H}_{B_1}\otimes \cdots \otimes {\cal H}_{B_{n}}$, the concurrence satisfies the following inequality
\begin{eqnarray}
C^2(|\psi_{\bf {A|B}})\leq (\sum^m_{i=1}C^2|\psi\rangle_{A_i|\overline{A_i}}))
\label{eqntg2}
\end{eqnarray}
where $\overline{A_i}=A_1\cdots A_{i-1}A_{i+1}\cdots A_mB_1\cdots B_n$.

{\bf Proof}. Combining Eq.(\ref{eqn01}) with Eq.(\ref{eqnLinear2}), we get
\begin{eqnarray}
C^2(|\psi\rangle_{\bf {A|B}})
&=&2T({\bf {A}})
\nonumber\\
&\leq& 2\sum^m _{i=1}T(A_i)
\nonumber
\\
&=&\sum^m_{i=1}C^2(\psi_{A_i|\overline{A_i}})
\label{}
\end{eqnarray}
for any $m+n$-qubit pure state $|\psi\rangle_{\bf{AB}}$ on Hilbert space ${\cal H}_{A_1}\otimes \cdots \otimes {\cal H}_{A_m}\otimes{\cal H}_{B_1}\otimes \cdots \otimes {\cal H}_{B_{n}}$. This completes the proof. $\hfill\square$

For later use we give the following lemma 2.

\textbf{Lemma 2} For any $a_i$ with $a_1\geq a_2\geq\cdots a_n\geq 0$, and $0\leq\mu\leq1$, the following inequality holds
\begin{eqnarray}
(\sum_{i=1}^na_i)^\mu&\leq& \sum_{i=0}^{n-1}((i+1)^\mu-i^\mu)a_{i+1}^\mu
\label{eqnlemma2}
\end{eqnarray}

The proof of Lemma 2 is firstly presented in ref.\cite{GYG19}. From Lemma 2, a similar inequality with a tightened upper bound than its shown in Theorem 7 is shown as follows.

\textbf{Theorem 8}. For any $m+n$-qubit pure state $|\psi\rangle_{\bf{AB}}$ on Hilbert space ${\cal H}_{A_1}\otimes \cdots \otimes {\cal H}_{A_m}\otimes{\cal H}_{B_1}\otimes \cdots \otimes {\cal H}_{B_{n}}$, assume that the concurrence satisfies $C(|\psi\rangle_{A_i|{\overline{A_i}}})\geq C(|\psi\rangle_{A_{i+1}|{\overline{A_{i+1}}}})$. Then we have
\begin{eqnarray}
C^\alpha(|\psi\rangle_{\bf {A|B}})\leq \sum^m_{i=1}(i^t-(i-1)^t)C^\alpha_{A_i|\overline{A_i}}
\label{eqntg3}
\end{eqnarray}
for all $0\leq\alpha\leq 2$ and $t=\alpha/2$.

{\bf Proof}. For any $m+n$-qubit pure state $|\psi\rangle_{\bf{AB}}$ on Hilbert space ${\cal H}_{A_1}\otimes \cdots \otimes {\cal H}_{A_m}\otimes{\cal H}_{B_1}\otimes \cdots \otimes {\cal H}_{B_{n}}$, from the inequality (\ref{eqntg2}), it is easy to deduce that
\begin{eqnarray}
C^\alpha(|\psi\rangle_{\bf {A|B}})&\leq& (\sum^m_{i=1}C^2|\psi\rangle_{A_i|\overline{A_i}})^t
\nonumber\\
&\leq&
\sum^m_{i=1}(i^t-(i-1)^t)C^\alpha_{A_i|\overline{A_i}}
\label{eqntg4}
\end{eqnarray}
Here, the first inequality is due to the inequality: $C(|\psi\rangle_{A_i|{\overline{A_i}}})\geq C(|\psi\rangle_{A_{i+1}|{\overline{A_{i+1}}}})$ for $i=1, 2, \cdots,  m-1$,  which implies the inequality of $C^2(|\psi\rangle_{A_i|{\overline{A_i}}})\geq C^2(|\psi\rangle_{A_{i+1}|{\overline{A_{i+1}}}})$. The second inequality is obtained from Lemma 2. $\hfill\square$

Note that Unified-$(q, s)$ entanglement contains EOF, Tsallis-$q$ and R\'{e}nyi-$q$ entanglement as special cases. For the Unified-$(q, s)$ entanglement measure, we prove interesting inequalities similar to the inequality (\ref{eqntg3}).

\textbf{Theorem 9}. For any $m+n$-qubit pure state $|\psi\rangle_{\bf{AB}}$  on Hilbert space ${\cal H}_{A_1}\otimes \cdots \otimes {\cal H}_{A_m}\otimes{\cal H}_{B_1}\otimes \cdots \otimes {\cal H}_{B_{n}}$, assume that Unified-$(q, s)$ entanglement satisfies $U_{q,s}(|\psi\rangle_{A_i|{\overline{A_i}}})\geq U_{q,s}(|\psi\rangle_{A_{i+1}|{\overline{A_{i+1}}}})$. Then we have
\begin{eqnarray}
U_{q,s}^\alpha(|\psi\rangle_{\bf {A|B}})\leq \sum^m_{i=1}(i^\alpha-(i-1)^\alpha)U_{q,s}^\alpha
(|\psi\rangle_{A_i|\overline{A_i}})
\label{eqntg5}
\end{eqnarray}
for all $0\leq\alpha\leq1$, where $q>1$ and $qs\geq1$.

{\bf Proof}. Denote $0\leq\alpha\leq1$, and $q>1$, $qs\geq1$, we have
\begin{eqnarray}
U_{q,s}(|\psi\rangle_{\bf {A|B}})&=&S_{q,s}(\rho_{\bf{A}})
\nonumber\\
&\leq&\sum^m_{i=1}S_{q,s}(\rho_{A_i})
\nonumber\\
&=&\sum^m_{i=1}U_{q,s}(|\psi\rangle_{A_i|\overline{A_i}})
\nonumber\\
&\leq&\sum^m_{i=1}(i^\alpha-(i-1)^\alpha)
U_{q,s}^\alpha(|\psi\rangle_{A_i|\overline{A_i}})
\label{}
\end{eqnarray}
Here, the first and second equalities are due to the definition of Unified-$(q, s)$ entanglement in Eq.(\ref{eqnUP}). The first inequality is from the fact that Unified entropy has the subadditivity
properties similar to the linear entropy as $S_{q,s}(\rho_{AB})\leq S_{q,s}(\rho_{A})+S_{q,s}(\rho_{B})$ for $q>1, qs\geq1$ \cite{unified entropies(2011)}. Moreover, from Lemma 2, we can deduce the last equality. $\hfill\square$

\section{Typical value of the tangle for high-dimensional states}

We have known that the aforementioned entanglement measure $E^{\alpha_c}$ is monogamous for qubit systems. A nature problem is whether the monogamy inequality (\ref{eqn1}) can  be generalized to higher dimensional systems or not.  In fact, Ou \cite{Ou(2007)} had indicated that the square of the concurrence (SC) is not monogamous for higher-dimension apart from qubit systems. In high-dimension case, another measure which is closely related to the concurrence, is named as the tangle that is an elementary entanglement measure \cite{tangle1(2003),tangle2(2010)}.

The tangle $\tau(|\psi\rangle)$ for
pure state $|\psi\rangle$ is defined by $\tau(|\psi\rangle)=C^2(|\psi\rangle)$, and the tangle for mixed state $\rho$ is defined by
\begin{eqnarray}
\tau'(\rho)=\inf_{\{p_i,|\phi_i\rangle\}}
\sum_ip_iM(\rho_{\varphi_i})
\end{eqnarray}
where $\rho=\sum_ip_i|\varphi_i\rangle\langle\varphi_i|$, $p_i\geq0$, and $\sum_ip_i=1$, $|\varphi_i\rangle\in \mathbb{C}^d\otimes \mathbb{C}^d$, $M(\rho)=2(1-\rm{Tr}(\rho))$.

Although the tangle and the concurrence are equivalent
as entanglement measures for pure states, they are different for mixed states. In fact, it holds
that $\tau'(\rho)\geq C^2(\rho)$ and the equality holds for two-qubit states \cite{T. J. Osborne(2005)}. Remarkably, it is a difficult to calculate the tangle of high-dimension mixed state because of a convex roof extension of the tangle. Fortunately, we can gain the typical value of $\tau'(\rho_{AB})$ of $\rho_{AB}$, a random state on $\mathbb{C}^d\otimes \mathbb{C}^d$ induced by an environment $\mathbb{C}^s$.

{\it Definition 3}. Given $n, s\in \mathbb{N}$, a random mixed state $\rho$ on $\mathbb{C}^n$
is induced by $\mathbb{C}^s$
if $\rho={\rm {Tr}}_{\mathbb{C}^s}(|\psi\rangle\langle\psi|)$ for $|\psi\rangle$, where $|\psi\rangle$ is
a uniformly distributed random pure state on  $\mathbb{C}^n\otimes \mathbb{C}^s$ \cite{Sara(2016)}.

In Sec.III, we have proved that the mentioned entanglement measures such as EOF and concurrence or its proper power of any entangled qubit states satisfy the monogamy inequality (1). However, this is impossible for for high-dimensional system. Informally, we show that the generic deviation between the typical value of the tangle and its average for a random induced state violate the monogamy inequality (1). Formally, we prove the typical value of the tangle as follows.

\textbf{Theorem 10.} Fix $t>0$. Let $\rho$ be a random state on $\mathbb{C}^d\otimes \mathbb{C}^d$ induced by some environment (auxiliary Hilbert) $\mathbb{C}^s$, with $s\leq Cd^2t^2$ for some universal constant $C>0$. Then,
\begin{eqnarray}
Pr(|\tau'(\rho)-(2-\frac{2d}{1+d^2})|\leq t)> 1-e^{-cd^2t^2}
\end{eqnarray}
where $c>0$ is a universal constant, $Pr(\cdot)$ is the probability function of the deviation between the typical value of the tangle for random induced states and its average.

The following crucial Lemmas will be used in order to establish the above Theorem 10.

\textbf{Lemma 3} \cite{Guillaume(2011),Sara(2016)}. Let $n\in \mathbb{N}$. For any $L$-Lipschitz function $g$: $S_{\mathbb{C}^n}\mapsto\mathbb{R}$
and any $t>0$, if $\psi$ is uniformly distributed on $S_{\mathbb{C}^n}$, then
\begin{eqnarray}
Pr(|g(\psi)-\mathbb{E }g|> t)\leq e^{-cnt^2/L^2}
\end{eqnarray}
where $c>0$ is a universal constant, $S_{\mathbb{C}^n}$ denotes the unit sphere $S$ of $\mathbb{C}^n$ and $\mathbb{E }X$ refers to the expectation of the random variable $X$.

\textbf{Lemma 4} \cite{Guillaume(2011),Sara(2016)}. Let $n\in \mathbb{N}$. For any circled $L$-Lipschitz function $g$: $S_{\mathbb{C}^n}\mapsto\mathbb{R}$
and any $t>0$, if ${\cal H}$ is a uniformly distributed $Cnt^2/L^2$-dimensional subspace of $\mathbb{C}^n$ with $C>0$ a universal constant, then
\begin{eqnarray}
Pr(\exists \psi\in {\cal H}\cap S_{\mathbb{C}^n}: |g(\psi)-\mathbb{E }g|> t)\leq e^{-cnt^2/L^2}
\end{eqnarray}
where $c>0$ is a universal constant.

\textbf{Lemma 5}. For $|\psi\rangle \in {\cal H}_A \otimes {\cal H}_B$, the Lipschitz constant $\eta$ (with respect to the Hilbert-Schmidt norm $\|\cdot\|_2$) of the function $g(|\psi\rangle)=2(1-\rm{Tr}(\rho^2_A))$
is upper bounded by 8, where $\rho_A$ denotes the reduced density operator of the subsystem $A$.

{\bf Proof}. Let $f(|\psi\rangle)=\sqrt{\rm{Tr}(\psi^2_A)}$, then $g(|\psi\rangle)=2-2f^2(|\psi\rangle)$. We have known from ref.\cite{Hayden(2006)} that the Lipschitz constant of the function $f(|\psi\rangle)$ is upper bounded by 2.
An elementary calculation yields
\begin{eqnarray}
\eta^2&=&\sup\nabla g\cdot\nabla g
\nonumber\\
&=&16\sup f^2\cdot(\nabla f)^2\nonumber\\
&\leq &16\sup (\nabla f)^2\nonumber\\
&\leq &64
\end{eqnarray}
where $\nabla g$ denotes the gradient of the function $g$. The Lipschitz constant $\eta$ of $g(|\psi\rangle)$ is then  bounded by 8.  $\hfill\square$

\textbf{Lemma 6}. Fix $n, s\in \mathbb{N}$ with $s\leq n$ and $t>0$. For a state $\psi$ being chosen uniformly from the unit sphere $S_{\mathbb{C}^{n}\otimes\mathbb{C}^{s}}$ on Hilbert space $\mathbb{C}^{n}\otimes\mathbb{C}^{s}$, we get
\begin{eqnarray}
Pr(|M(\rho_{\psi})-(2-\frac{2(n+s)}{1+ns})|> t)\leq e^{-cnst^2}
\end{eqnarray}
where $c>0$ is a universal constant.

{\bf Proof}. Define the function $g: \psi\mapsto M(\rho_{\psi})$, $\forall \psi\in S\,_{\mathbb{C}^{n}\otimes\mathbb{C}^{s}}$ and $M(\rho_{\psi})$ is defined in Eq.(66). Lemma 5 reveals that $g$ is 8-Lipschitz.  We know from ref.\cite{Lubkin(1978)}, that $g$ has average $\mathbb{E } g(\psi)=2-\frac{2(n+s)}{1+ns}$. Using these results, Lemma 6 is a direct consequence of Lemma 3. $\hfill\square$

\textbf{Lemma 7}. Fix $d\in \mathbb{N}$  and $t>0$. Suppose that ${\cal H}$ is a uniformly distributed  $Cd^2t^2$-dimensional subspace of $\mathbb{C}^{d}\otimes\mathbb{C}^{d}$, $C>0$ is a universal constant. And then $\exists \psi\in {\cal H}\cap{S\,_{\mathbb{C}^{d}\otimes\mathbb{C}^{d}}}$ satisfies \begin{eqnarray}
Pr( |M(\rho_{\psi})-(2-\frac{2d}{1+d^2})|> t)\leq e^{-cd^2t^2}
\end{eqnarray}

{\bf Proof}. Define the function $g: \psi\mapsto M(\rho_{\psi})$, $\forall \psi\in S\,_{\mathbb{C}^{d}\otimes\mathbb{C}^{d}}$. $g$ is 8-Lipschitz from Lemma 5. Furthermore, $g$ has average $\mathbb{E }g(\psi)=2-\frac{2d}{1+d^2}$ from ref.\cite{Lubkin(1978)}. These two results implies Lemma 7 according to Lemma 4.  $\hfill\square$

\textbf{Lemma 8}. Fix $t>0$. Let $\rho$ be a random state on $\mathbb{C}^d\otimes \mathbb{C}^d$ induced by some environment (axillary space) $\mathbb{C}^s$ with $s\leq Cd^2t^2$ for some universal constant $C>0$. Then, $\exists \psi\in {\rm{supp}}(\rho)\cap S_{\mathbb{C}^{d}\otimes\mathbb{C}^{d}}$ satisfies
\begin{eqnarray}
Pr(|M(\rho_{\psi})-(2-\frac{2d}{1+d^2})|\leq t)>1-e^{-cd^2t^2}
\end{eqnarray}
where ${\rm{supp}}(\rho)$ denotes the support of the density operator $\rho$.

{\bf Proof}. From the assumption of $\rho$, the support of the density operator, i.e.,  ${\rm{supp}}(\rho)$ is a subspace of $\mathbb{C}^d\otimes \mathbb{C}^d$. By using Lemma 7 we obtain the result.  $\hfill\square$

From Lemma 8 and Eq.(66), we can obtain Theorem 10.

\section{Examples}

In this section, we present some entangled states to show the generalised monogamy inequalities of $\alpha$-th power of bipartite entanglement measure for $\alpha\geq \alpha_{c}$. The following examples include generalized tripartite system \cite{Acin(2000)}, decoherence-free state \cite{Kempe(2001),Zhou(2016)}, Dicke state \cite{Karmakar(2016)}. As its shown in Sec.II, we can evaluate all the mentioned entanglement measures from the concurrence. Thus, we do not show EOF, Tsallis-$q$ entropy,  R\'{e}nyi-$q$ entropy and Unified-$(q, s)$ entropy.

{\it Definition 4}. For a multipartite quantum state $\rho_{A{\bf B}}$, define the residual quantity of $\alpha$-th power of bipartite entanglement measure $E$ as
\begin{eqnarray}
 \tau_E(\rho_{A{\bf B}})
&=&E^\alpha(\rho_{A|{\bf B}})
-\sum_{j=0}^{n-1} ((\gamma_0+1)^t-\gamma_0^t)^{\omega_H(\vec{j})}
\nonumber\\
&&\times E^\alpha(\rho_{A|B_j})
\label{eqn288}
\end{eqnarray}
where $t=\alpha/{\alpha_{c}}$, $\gamma_0=\gamma^{\alpha_{c}}$, and $\omega_H(\vec{j})$ denotes the Hamming weight of vector $\vec{j}$.

The value of $\alpha_c$ depends on the specific measure $E$, i.e., $\alpha_c=2$ for the concurrence (Tsallis-$q$ entropy , R\'{e}nyi-$q$ entropy and Unified-$(q, s)$ entropy entanglement), and $\alpha_c=\sqrt{2}$ for the EOF.

\subsection{Three-qubit entangled state}

Under local unitary operations, any three-qubit entangled pure state can be changed into the following normal form as \cite{Acin(2000)}:
\begin{eqnarray}
|\psi\rangle
&=&\lambda_0|000\rangle+\lambda_1e^{i\varphi}|100\rangle+\lambda_2|101\rangle
\nonumber\\
&&+\lambda_3|110\rangle
+\lambda_4|111\rangle
\label{eqn29}
\end{eqnarray}
where $\lambda_i\geq0$, $0\leq\varphi\leq \pi$, and $\sum_{i=0}^4\lambda_i^2=1$. For convenience of discussing the monogamy inequality in Theorem 1, assume that two parameters of $|\psi\rangle_{ABC}$ are given by: $\lambda_2=\frac{\sqrt{2}}{2}, \lambda_3=\frac{1}{2}$.
Now, define parametric representations of $\lambda_i$s as
$\lambda_0=\frac{1}{2}\sin\phi\cos\theta$,
$\lambda_1=\frac{1}{2}\sin\phi\sin\theta$ and
$\lambda_4=\frac{1}{2}\cos\phi$ for $0\leq\phi\leq \pi/2$ and
$0\leq\theta\leq \pi/2$. The concurrence of $|\psi\rangle_{ABC}$ is given by
\begin{eqnarray}
&&C(|\psi\rangle_{A|BC})=2\lambda_0\sqrt{1-\lambda_0^2-\lambda_1^2}
\nonumber\\
&&C(\rho_{A|B})=\lambda_0
\nonumber\\
&&C(\rho_{A|C})=\sqrt{2}\lambda_0
\label{eqn31}
\end{eqnarray}
It follows that $C(\rho_{A|C})\geq \sqrt{2} C(\rho_{A|B})$. The residual quantity of the concurrence $\tau_{C}(|\psi\rangle)$ is
\begin{eqnarray}
\tau_{C}(|\psi\rangle)&=&
C^\alpha(|\psi\rangle_{A|BC})-C^\alpha(\rho_{A|C})
\nonumber\\
&&-(3^{\alpha/2}-2^{\alpha/2})C^\alpha(\rho_{A|B})
\label{eqn311}
\end{eqnarray}
Combined with Eqs.(\ref{eqn31}) and (\ref{eqn311}), the numeric evaluations of the residual quantity of all  entanglement measures are shown in Fig.1. It is easy to check that $\tau_{C}(|\psi\rangle)$, $\tau_{E_f}(|\psi\rangle)$, $\tau_{T_q}(|\psi\rangle)$,  $\tau_{R_q}(|\psi\rangle)$, and $\tau_{U_{q,s}}(|\psi\rangle)$ are always positive for $0\leq\theta\leq \pi/2$,
and $0\leq\phi\leq \pi/2$. Namely, the inequality (\ref{eqn6}) holds for three-qubit entangled pure state in terms of the entanglement measures derived from the concurrence, the EOF, Tsallis-$q$ entropy, R\'{e}nyi-$q$ entropy, and Unified-$(q, s)$ entropy.

\begin{figure}
\begin{center}
\resizebox{260pt}{210pt}{\includegraphics{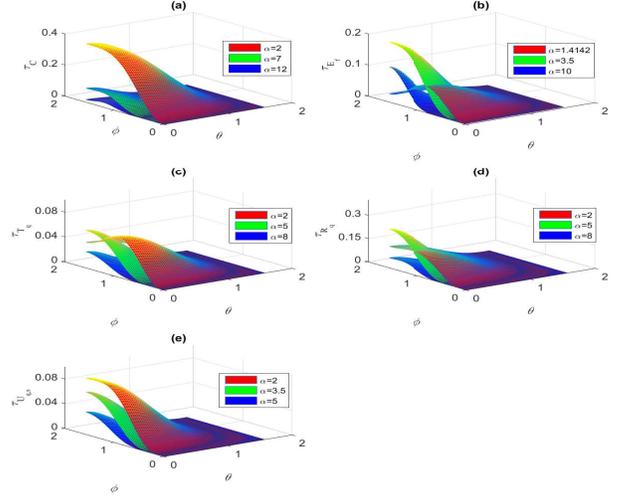}}
\end{center}
\caption{\small (Color online) {\bf The residual quantity $\tau$ for the three-qubit state $|\psi\rangle$}. (a) $\tau_{C}$ of the concurrence. (b) $\tau_{E_f}$ of EOF. (c) $\tau_{T_q}$ of Tsallis-$q$ entropy with $q=0.7$. (d) $\tau_{R_q}$ of  R\'{e}nyi-$q$ entropy with $q=0.9$. (e) $\tau_{U_{q,s}}$ of Unified $(q,s)$ entropy with $q=1.1$ and $s=0.4$. Here, $0\leq\theta\leq \pi/2$ and $0\leq\phi\leq \pi/2$.
}
\label{fig1}
\end{figure}

\subsection{Decoherence-free state}

In applications, it is inevitable for decoherence being induced by uncontrolled coupling between a quantum system and the environment. When qubits are coupled to the environment, the quantum superposition and coherence are easily destructed, and as a result the maximally entangled state collapses into a non-maximally entangled one or even a mixed state. This will degrade the fidelity and security of quantum communication. To overcome this flaw, specific entangled states, named as decoherence-free
states \cite{Kempe(2001)}, are proposed. Decoherence-free states, no matter how strong the qubit-environment interaction, exhibit some symmetry, which are useful for long-distance quantum information transmission and storage.

A four-qubit entangled decoherence-free state is given by \cite{Zhou(2016)}:
\begin{eqnarray}
|\Phi\rangle=a|\Psi_0\rangle+b|\Psi_1\rangle
\label{eqnfreestate}
\end{eqnarray}
where $|\Psi_i\rangle$ are logic basis states given by
\begin{eqnarray}
|\Psi_0\rangle_{ABCD}&=&\frac{1}{2}(|01\rangle-|10\rangle)_{AB}(|01\rangle-|10\rangle)_{CD},
\nonumber\\
|\Psi_1\rangle_{ABCD}&=&\frac{1}{2\sqrt{3}}(2|1100\rangle+
2|0011\rangle-|1010\rangle-|1001\rangle
\nonumber\\
&&-|0101\rangle-|0110\rangle)_{ABCD}
\label{11}
\end{eqnarray}
For the purpose of discussing monogamy inequality (\ref{eqn14}) entanglement, we take three cases into account for decoherence-free state in Eq.(\ref{eqnfreestate}).

When $a=b=\frac{1}{\sqrt{2}}$
the concurrence for $|\Phi\rangle$ are computed as
$C(|\Phi\rangle_{A|BCD})=1$,  $C(\rho_{A|B})=0.9107$,
$C(\rho_{A|C})=0.3333$ and $C(\rho_{A|D})=0.244$. We get $\gamma=1.3$. The residual quantity of the concurrence
is given by
\begin{eqnarray}
\tau_{C}(|\Phi\rangle)
&=&C^\alpha(|\Phi\rangle_{A|BCD})-C^\alpha(\rho_{A|B})
\nonumber\\
&&-((\gamma^2+1)^{\alpha/2}-\gamma^{\alpha})C^\alpha(\rho_{A|C})
\nonumber\\
&&-((\gamma^2+1)^{\alpha/2}-\gamma^{\alpha})C^\alpha(\rho_{A|D})
\label{eqn39}
\end{eqnarray}
Similarly, we can obtain entanglement measures for different $a, b$. The residual quantities of
all measurement measures for decoherence-free state $|\Phi\rangle$ are shown in Fig.2. The residual quantity of the entanglement measures are always positive, which indicates that decoherence-free state is monogamous for the $\alpha$-th power entanglement measures with $\alpha\geq \alpha_c$.

\begin{figure}
\begin{center}
\resizebox{260pt}{210pt}{\includegraphics{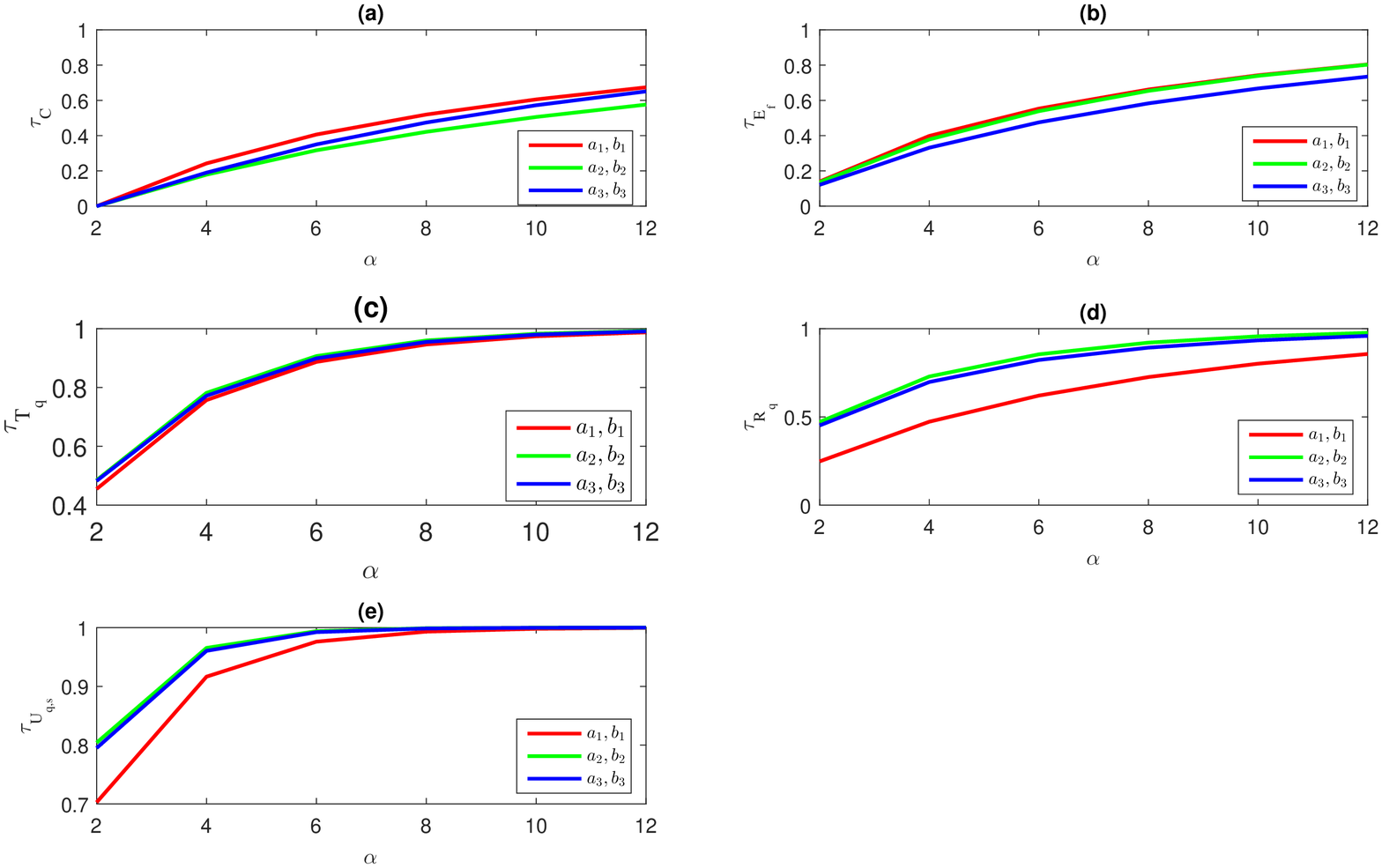}}
\end{center}
\caption{\small (Color online) {\bf The residual quantity $\tau$ for decoherence-free state $|\Phi\rangle$}. (a) $\tau_{C}$ of the concurrence. (b) $\tau_{E_f}$ of EOF. (c) $\tau_{T_q}$ of Tsallis-$q$ entropy with $q=0.7$. (d) $\tau_{R_q}$ of  R\'{e}nyi-$q$ entropy with $q=1.2$. (e) $\tau_{U_{q,s}}$ of Unified $(q,s)$ entropy with $q=1.4$ and $s=0.6$. Here, $a_1=\frac{1}{\sqrt{2}}$, $a_2=\frac{\sqrt{6}}{3}$, $a_3=\frac{2}{\sqrt{5}}$, $b_1=\frac{1}{\sqrt{2}}$, $b_2=\frac{1}{\sqrt{3}}$, and $b_3=\frac{1}{\sqrt{5}}$.}
\end{figure}

\indent
\subsection{Dicke state}

The symmetric systems are experimentally interesting because it is easier to nonselectively address an entire ensemble of particles rather than individually address each member \cite{Stockton(2003)}. An $n$-qubit Dicke state \cite{Karmakar(2016)} with $k$ excitations is given by
\begin{eqnarray}
|D^{(k)}_n\rangle_{A_1A_2\cdots A_n}=\frac{1}{\sqrt{\binom{n}{k}}}\sum_{perm}(|0\rangle^{\otimes (n-k)}|1\rangle^{\otimes k})
\end{eqnarray}
where the summation is over all possible permutations of the product states having $n-k$ zeros and $k$ ones, and $\binom{n}{k}$ denote the combination number choosing $k$ items from $n$ items.

The concurrences for Dicke state are given by
\begin{eqnarray}
C(|D^{(k)}_n\rangle_{A_1|A_2\cdots A_n})&=&\frac{2\sqrt{k(n-k)}}{n}
\nonumber\\
C(|D^{(k)}_n\rangle_{A_1|A_i})&=&
-\frac{2\sqrt{k(k-1)(n-k)(n-k-1)}}{n(n-1)}
\nonumber
\\
&&+ \frac{2k(n-k)}{n(n-1)}
\label{}
\end{eqnarray}
where, $i\in\{2, \cdots, n\}$.  Here, consider three cases with $n=4, n=5$ and $n=6$. We get
\begin{eqnarray}
&&C(|D^{(2)}_4\rangle_{A_1|A_2A_3A_4})=1
\nonumber\\
&&C(|D^{(2)}_4\rangle_{A_1|A_i})=\frac{1}{3}, i\in\{2, 3, 4\}
\nonumber\\
&&
C(|D^{(2)}_5\rangle_{A_1|A_2\cdots{}A_5})=\frac{2\sqrt{6}}{5} \nonumber\\
&&C(|D^{(2)}_5\rangle_{A_1|A_i})=\frac{3-\sqrt{3}}{5}, i\in\{2, \cdots, 5\}
\nonumber\\
&&C(|D^{(3)}_6\rangle_{A_1|A_2\cdots{}A_6})=1
\nonumber\\
&&C(|D^{(3)}_6\rangle_{A_1|A_i})=\frac{1}{5}, i\in\{2, \cdots, 6\}
\label{eqnDicke3}
\end{eqnarray}
Thus, the residual quantity of the concurrence for Dicke state $|D^{(2)}_4\rangle$, $|D^{(2)}_5\rangle$, $|D^{(3)}_6\rangle$ have the following forms, respectively
\begin{eqnarray}
\tau_{C}(|D^{(2)}_4\rangle)
&=&1-\frac{1}{3^\alpha}-2(2^{\alpha/2}-1)\frac{1}{3^\alpha}
\nonumber\\
\tau_{C}(|D^{(2)}_5\rangle)
&=&(\frac{2\sqrt{6}}{5})^\alpha
-2^{\alpha}(\frac{3-\sqrt{3}}{5})^\alpha
\nonumber\\
\tau_{C}(|D^{(3)}_6\rangle)
&=&1-\frac{1}{5^\alpha}-
\frac{2^{\alpha}+2^{\alpha/2}-2}{5^\alpha}
\label{}
\end{eqnarray}
which are shown in Fig.3(a) for $0\leq\alpha\leq 5$. Similarly, we obtain the residual quantities in terms of different measures, as showed in Fig.3. It shows that inequality (\ref{eqn14}) holds for the $\alpha$-th power of entanglement measures with $0\leq\alpha\leq 5$ with respect to Dicke state, in fact, one can find that it still works for $\alpha\geq\alpha_c$.

\begin{figure}
\begin{center}
\resizebox{260pt}{210pt}{\includegraphics{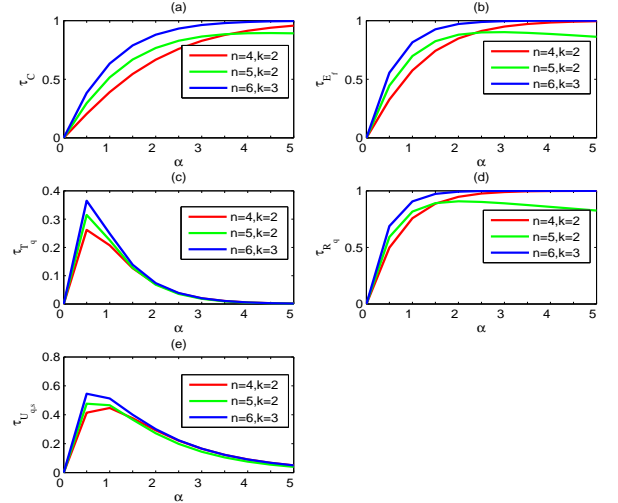}}
\end{center}
\caption{\small (Color online) {\bf The residual quantity $\tau$ for Dicke state $|D^{(2)}_4\rangle$ (the red line), $|D^{(2)}_5\rangle$ (green line), and $|D^{(3)}_6\rangle$ (blue line)}. (a) $\tau_{C}$ of the concurrence. (b) $\tau_{E_f}$ of EOF. (c) $\tau_{T_q}$ of Tsallis-$q$ entropy with $q=4.3$. (d) $\tau_{R_q}$ of  R\'{e}nyi-$q$ entropy with $q=1.3$. (e) $\tau_{U_{q,s}}$ of Unified $(q,s)$ entropy with $q=2$ and $s=0.7$.}
\end{figure}

\section{Conclusion}

Given an entangled state, how much is it entangled? To address this problem, the concept of entanglement measure has been naturally arisen. One intrinsic feature of quantum entanglement is the limited shareability of bipartite entanglement in multipartite quantum systems. This distinct property without any classical counterpart is known as the monogamy of entanglement (MOE). In this paper, we investigated the monogamy property of $\alpha$-th power of entanglement measures in bipartite entangled qubit states. So far, there are several well-known bipartite entanglement measures which quantify the degree of entanglement, we focus on the unified monogamy relations of bipartite entanglement measures. We establish a class of weighted monogamy inequalities of multipartite entangled qubit systems based on the unified monogamy inequality. Moreover, we shown that the present monogamy inequalities are tighter than previous results. Additionally, some generalized monogamy inequalities related to bipartite entanglement measures are obtained. For high-dimensional entangled states, we proved a different result, which shows that the generic deviation between the typical value of the tangle and its average for a random induced state violate the monogamy inequality (1).  These results are interesting in the entanglement theory, quantum information processing, quantum communication, and quantum many-body theory.

\section*{Acknowledgments}

This work was supported by the National Natural Science Foundation of China (No.61772437), Sichuan Youth Science and Technique Foundation (No.2017JQ0048), and Fundamental Research Funds for the Central Universities (No.2018GF07).


\end{document}